\documentclass[%
 reprint,
 longbibliography,
 amsmath,amssymb,
 aps,
]{revtex4-2}
\usepackage{graphicx}
\usepackage{dcolumn}
\usepackage{bm}
\usepackage{gensymb}
\usepackage{braket}  
\usepackage[colorlinks=true, linkcolor=blue, citecolor=blue, urlcolor=blue]{hyperref}
\usepackage[markup=default]{changes}
\definechangesauthor[name={Wang}, color=orange]{W}

\begin{document}

\preprint{APS/123-QED}

\title{Spectral-Domain Coherent Control of Broadband Raman Coupling \\ in Atom Interferometry}

\author{Sheng-Zhe Wang}
\author{Wei-Chen Jia}
\author{Yue Xin}
\author{Qian-Lan Cai}
\author{Yingpeng Zhao}
\author{Yan-Ying Feng \(\)}

\email{yyfeng@tsinghua.edu.cn}

\affiliation{
\(^{1}\) A-Knows Lab, State Key Laboratory of Precision Measurement Technology and Instruments, Department of Precision Instruments, Tsinghua University, Beijing 100084, China}


\date{\today}
             
\begin{abstract}
The performance of atom interferometers is commonly limited by the finite spectral acceptance of atomic beam splitters and mirrors, which restricts efficient coupling to atoms with large Doppler shifts and reduces the usable atomic flux. Here, we demonstrate spectral-domain coherent control of Raman coupling by engineering its effective two-photon spectrum. By synthesizing multiple frequency components, the Raman
interaction simultaneously addresses a broad range of atomic velocities, effectively overcoming the conventional transit-time-limited linewidth. Implemented in a continuous atomic-beam Mach-Zehnder interferometer, where the transverse Doppler broadening is 17 times larger than the intrinsic Raman linewidth, this approach enhances the fringe contrast from $5.9(2)\%$ to $15.1(2)\%$, indicating a substantial 
increase in effective atomic participation. Our results establish spectral-domain coherent control as a general strategy for achieving spectrally robust atom interferometry and open new opportunities for quantum sensing in systems with strong inhomogeneous broadening.
\end{abstract}

\maketitle

Atom interferometers have become powerful tools for a wide range of applications, including inertial sensing \cite{Gustavson1997,Canuel2006,Dickerson2013,Meng2024}, gravitational-wave detection \cite{Badurina2020,Abe2021}, dark matter searches \cite{Badurina2020,Hamilton2015}, and precision tests of fundamental physics \cite{Safronova2018,Asenbaum2020}. Their performance critically relies on high-fidelity coherent manipulation of atomic wave packets via atom-light interactions. However, the finite spectral acceptance of these interactions limits efficient coupling to atoms with large detunings. In many practical scenarios, the Doppler-detuning distribution arising from the atomic velocity spread is much broader than the transition linewidth, thereby reducing both effective atomic participation and interferometric contrast. Although laser cooling can mitigate Doppler broadening, achieving sufficiently narrow velocity distributions often requires additional time or space for preparation, introducing dead time and system complexity. Alternatively, the spectral response can be broadened by reducing the laser beam waist \cite{Kwolek2022} or shortening the pulse duration \cite{Kasevich1991}, however, these approaches are ultimately limited by wavefront-inhomogeneity-induced dephasing \cite{Schkolnik2015,Zhou2016,Xu2024} and constraints on achievable transient laser power \cite{Dutta2016}.

A variety of coherent-control techniques have been developed to improve manipulation fidelity, including adiabatic rapid passage \cite{Marte1991,Weitz1994,Kotru2015}, composite pulse sequences \cite{Butts2013,Dunning2014,Berg2015}, and quantum optimal control \cite{Saywell2018,Wilkason2022,Saywell2023}. These approaches enhance robustness against Doppler detuning by shaping the temporal profiles of the driving fields. However, they typically require longer interaction times, leading to increased spontaneous-emission loss. Moreover, implementing dynamic temporal control of pulse parameters is challenging in continuous platforms, such as continuous atom-beam interferometers \cite{Durfee2006,Xue2015,Kwolek2022,Meng2024,Sato2025,Yan2025}. These limitations highlight the need for an alternative paradigm based on spectral-domain coherent control.

In this Letter, we introduce a spectral-domain strategy that provides a fundamentally distinct paradigm for coherent control. Rather than modifying the temporal profile of the interaction, this approach directly engineers the spectral response while preserving the original pulse shapes. By synthesizing multiple effective two-photon frequencies, we broaden the spectral response of stimulated Raman transitions, enabling simultaneous addressing of atoms across a thermal velocity distribution. In other areas of atomic physics, related concepts have been used to spectrally broaden single-photon interactions for white-light cooling \cite{Hoffnagle1988,Zhu1991}, modulation transfer spectroscopy \cite{Guan2025}, and optical clocks \cite{Shang2022,Tobias2025}. These results suggest that extending coherent control into the spectral domain provides a natural route to overcoming Doppler-induced limitations.

\begin{figure*}[t]
\includegraphics[width=1\linewidth]{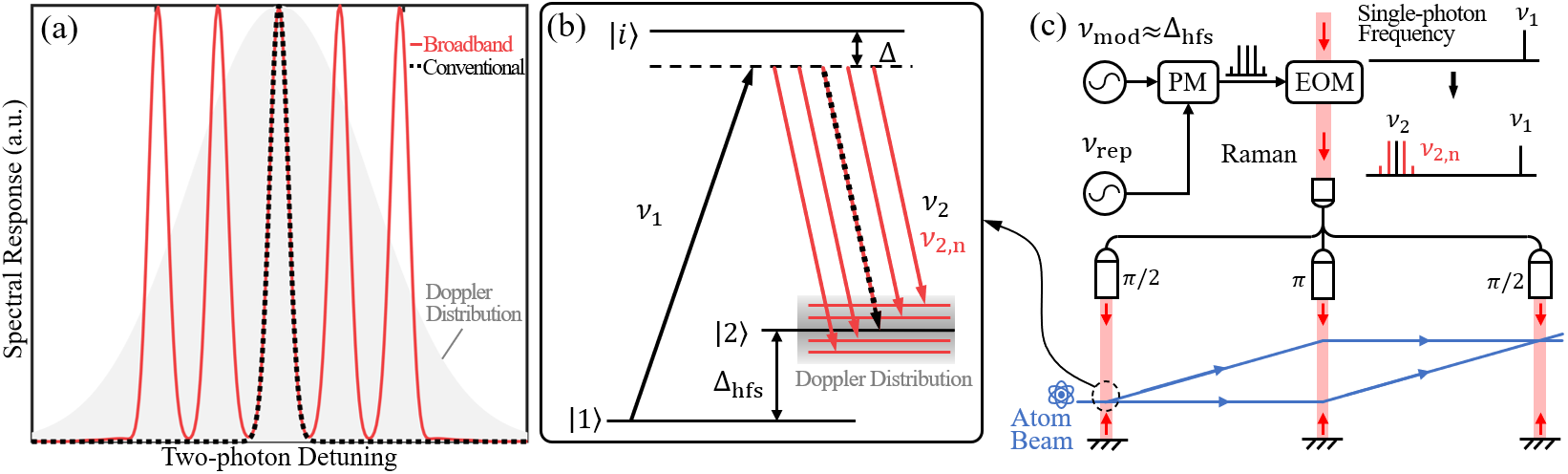}
\caption{\label{fig:Principle} Spectral-domain engineering of broadband Raman transitions. (a) Two-photon spectral response of conventional (black dashed) and broadband (red) Raman transitions, compared with the Doppler distribution of the atoms (gray). The broadband scheme synthesizes multiple resonances in two-photon detuning, enabling efficient coupling over a wide velocity range. (b) Level diagram of the Raman transition between $\ket{1}$ and $\ket{2}$ via the excited state $\ket{i}$ with single-photon detuning $\Delta$. Multiple two-photon resonances arise from the frequency components $\nu_{1}$ and $\nu_{2,n}$. (c) Experimental implementation in a continuous $^{87}$Rb atomic beam interferometer. Broadband Raman fields are generated using a phase modulator (PM) and an electro-optic modulator (EOM), enabling spectrally engineered coherent manipulation across the thermal ensemble.}
\end{figure*}

The principle of spectral-domain coherent control is illustrated in Fig.~\ref{fig:Principle}. We implement this approach in a Mach-Zehnder atom interferometer, where the broadened spectral response enhances the transfer efficiency of the beam splitters and mirrors across the thermal velocity distribution, thereby increasing both the fringe contrast and effective atomic participation. More generally, this spectral-domain strategy provides a general framework for robust coherent control in  quantum systems subject to inhomogeneous detunings.

In our approach, coherent control of stimulated Raman transitions is engineered in the two-photon spectral domain. Due to the Doppler effect, atoms are distributed in two-photon detuning as $\delta = k_{\rm eff} v$, where $v$ is the transverse velocity and $k_{\rm eff}$ is the effective Raman wave number \cite{Kasevich1991a}. In conventional Raman transitions, the spectral response is transit-time limited [black dashed in Fig.~\ref{fig:Principle}(a)] and typically much narrower than the Doppler distribution, so that only a small fraction of atoms are efficiently addressed. 

To quantify this limitation, we describe the Raman interaction by its spectral response $\rho(\delta)$, defined as the transfer probability as a function of two-photon detuning. Only atoms whose detunings fall within this response can be efficiently driven. The total transfer probability is therefore given by

\begin{equation}
P \propto \int \rho(\delta)\,f(\delta)\,d\delta,
\label{eq:P}
\end{equation}
which represents the spectral overlap between the Raman response and the Doppler distribution $f(\delta)$.

In the broadband scheme, the spectral response is engineered by introducing multiple two-photon resonances [red in Fig.~\ref{fig:Principle}(a)]. This effectively replaces a single narrow resonance with a comb of spectrally shifted responses,
\begin{equation}
\rho(\delta) \rightarrow \sum_n \rho(\delta - \delta_{L,n}),
\label{eq:rhoadd}
\end{equation}
thereby increasing the overlap with the Doppler distribution and enhancing the total transfer probability. Physically, atoms that are off-resonant for one component can be resonantly addressed by others, enabling efficient coupling across multiple velocity classes.

As shown in Fig.~\ref{fig:Principle}(b), these multiple resonances are realized using a Raman field composed of a single component at $\nu_1$ and multiple components at $\nu_{2,n}$. The electric field can be written as

\begin{equation}
\begin{aligned}
E = &\, E_{1}\cos(2\pi\nu_1 t+\phi_1) \\
    &+ E_{2}\sum_{n}\alpha_n \cos(2\pi\nu_{2,n} t+\phi_{2,n}),
\end{aligned}
\label{eq:E}
\end{equation}
where $\alpha_n$ and $\phi_{2,n}$ denote the relative amplitudes and phases. For large single-photon detuning $\Delta$, the excited state $\ket{i}$ can be adiabatically eliminated, yielding an effective two-level system with Hamiltonian
\begin{equation}
H_{\rm eff} =
\begin{pmatrix}
\delta & \Omega_{\rm eff,broad} \\
\Omega^*_{\rm eff,broad} & 0
\end{pmatrix},
\label{eq:H}
\end{equation}
where the effective coupling is given by
\begin{equation}
\Omega_{\rm eff,broad} = \sum_{n} \Omega_{\rm eff,n} e^{-i(\delta_{L,n} t + \phi_n)},
\label{eq:Omega}
\end{equation}
with $\Omega_{\rm eff,n} = \Omega_{2,n}^{*}\Omega_1/(2\Delta)$ and $\delta_{L,n} = 2\pi(\Delta_{\rm hfs} - \nu_1 + \nu_{2,n})$. The superposition of these phase-coherent couplings gives rise to the multi-resonance spectral response in Eq.~\ref{eq:rhoadd}.

Numerical solutions of Eq.~\ref{eq:H} yield the broadband Raman spectrum shown in Fig.~\ref{fig:Principle}(a). For equally spaced frequency components, the spectrum forms a comb of discrete resonances with spacing $\nu_{\rm rep}$. Each resonance behaves independently with a transit-time-limited linewidth $\delta\nu$, provided $\nu_{\rm rep} > \delta\nu$. By matching the induced detunings $\delta_{L,n}$ to the Doppler shifts $k_{\rm eff}v$, the interaction bandwidth is effectively extended across the atomic ensemble.

This spectral-domain approach provides a complementary paradigm to conventional temporal-domain control, where bandwidth is typically increased by temporal shaping. The two approaches thus establish a conceptual duality between the temporal shaping and spectral engineering of coherent interactions.


Experimentally, we realize phase-coherent multi-frequency Raman coupling by imprinting microwave sidebands onto the $\nu_2$ optical field using an electro-optic modulator (EOM), with an additional phase modulation at frequency $\nu_{\rm ref}$ [Fig.~\ref{fig:Principle}(c)]. This generates a set of frequency components $\nu_{2,n}$ while preserving a clean $\nu_1$ component. The amplitudes $\alpha_n$ follow the Bessel-function distribution $J_n(\beta)$ determined by the modulation depth $\beta$. This purely electro-optic implementation enables broadband spectral engineering without additional optical complexity, providing a simple and robust route to spectral-domain coherent control applicable to a wide range of quantum sensing platforms.

\begin{figure}[t]
\includegraphics[width=1\linewidth]{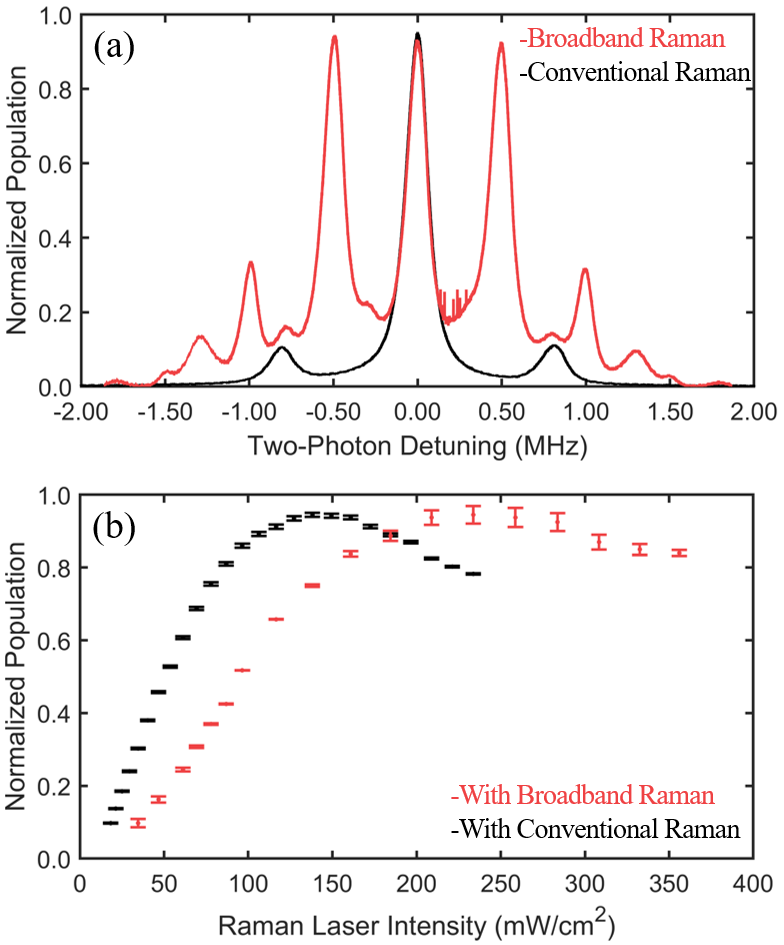}
\caption{\label{fig:ISRamanRabi}Spectral characterization of broadband Raman coupling in a Doppler-free configuration. (a) Raman spectra obtained with conventional (black) and broadband (red) Raman fields. The broadband scheme produces a comb of discrete resonances spaced by $\nu_{\rm rep}=500~\mathrm{kHz}$, forming an engineered spectral response that extends beyond the conventional linewidth. Minor peaks at $\pm 0.8~\mathrm{MHz}$ and $\pm 0.13~\mathrm{MHz}$ originate from transitions between magnetic sublevels. (b) Rabi oscillations as a function of Raman laser intensity, confirming coherent population transfer across multiple spectral components.}

\end{figure}

We demonstrate the spectral-domain coherent control in a thermal $^{87}$Rb atomic-beam interferometer following Ref.~\cite{Yan2025}. The atomic beam has a mean velocity of $175~{\rm m/s}$, and the Raman beams have a spatial width of $1~{\rm mm}$, corresponding to an effective pulse duration of $5.71~\mu{\rm s}$ and a transit-time-limited linewidth of $175~{\rm kHz}$ (FWHM). In contrast, the Doppler distribution of the thermal ensemble is approximately $3.0~{\rm MHz}$ (FWHM), an order of magnitude larger than the Raman linewidth. As a result, conventional Raman transitions can only efficiently address a small fraction of the atomic ensemble, leading to limited transfer efficiency and reduced interferometric contrast. The broadband scheme effectively extends the spectral acceptance of the Raman transition, increasing the number of addressed velocity classes. In this sense, the broadband scheme provides a spectral coverage comparable to that obtained with tightly focused Raman beams of about $100~\mu{\rm m}$ waist, while maintaining a well-collimated beam of $1~\mathrm{mm}$. This demonstrates that spectral-domain control relaxes the fundamental trade-off between spectral bandwidth and beam geometry inherent to temporal-domain approaches.

We first characterize the spectral response of the broadband Raman transition using a Doppler-free configuration, where the laser-induced two-photon detunings are swept synchronously to mimic the response of atoms at each given Doppler shift. As shown in Fig.~\ref{fig:ISRamanRabi}(a), the conventional Raman transition exhibits a single narrow resonance, whereas the broadband scheme produces a set of discrete resonances spaced by $\nu_{\rm rep}=500~{\rm kHz}$. These resonances correspond to multiple effective two-photon detunings and collectively form a broadened spectral response, in direct correspondence with the spectral-engineering picture illustrated in Fig.~\ref{fig:Principle}(a). The nearly equal amplitudes of the central resonances ($n=-1,0,1$) indicate uniform coupling strengths across multiple frequency components, enabling efficient addressing of atoms over an extended detuning range. This directly verifies the spectral-engineering picture of a multi-resonance comb that extends the effective interaction bandwidth.


Since the optical power is distributed among multiple components, the effective Rabi frequency is reduced by $\sqrt{N}$ for $N$ equal-strength resonances, consistent with $\Omega_{\rm eff} \propto \sqrt{I_1 I_2}$. As shown in Fig.~\ref{fig:ISRamanRabi}(b), a broadband $\pi$ pulse requires approximately $1.7$ times higher intensity, in agreement with the expected $\sqrt{3}$ scaling for three dominant resonances.

\begin{figure}[t]
\includegraphics[width=0.98\linewidth]{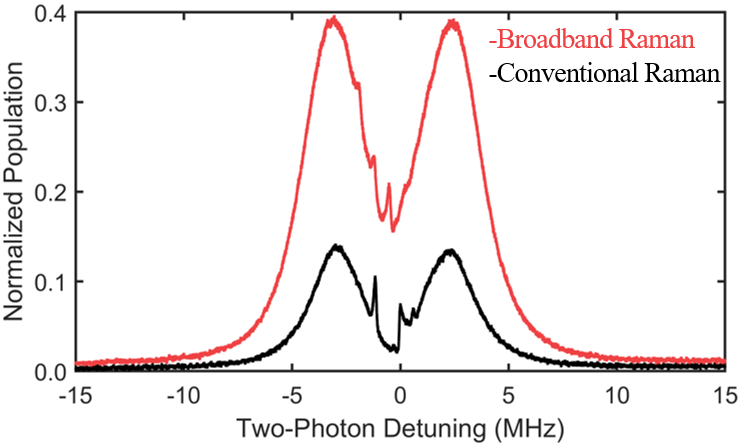}
\caption{\label{fig:DSRamanRabi}Doppler-sensitive Raman transition spectra in the thermal atomic beam. The broadband Raman scheme (red) significantly increases the transfer efficiency compared to the conventional case (black), demonstrating enhanced coupling across the Doppler-broadened velocity distribution.}
\end{figure}

While the Doppler-free spectrum reveals the engineered spectral structure, the Doppler-sensitive Raman transition directly probes the transfer efficiency across the thermal velocity distribution. As shown in Fig.~\ref{fig:DSRamanRabi}, the conventional Raman transition yields a population transfer of only $0.14(2)$, reflecting the limited fraction of atoms within the narrow resonance window. In contrast, the broadband scheme increases the transfer efficiency to $0.39(6)$, corresponding to an approximately threefold enhancement. This increase is consistent with the number of dominant resonances in the broadened spectrum and directly demonstrates the improved overlap between the Raman interaction and the atomic velocity distribution.This provides direct evidence that spectral-domain engineering increases the effective participation of the atomic ensemble.

\begin{figure}[t]
\includegraphics[width=1\linewidth]{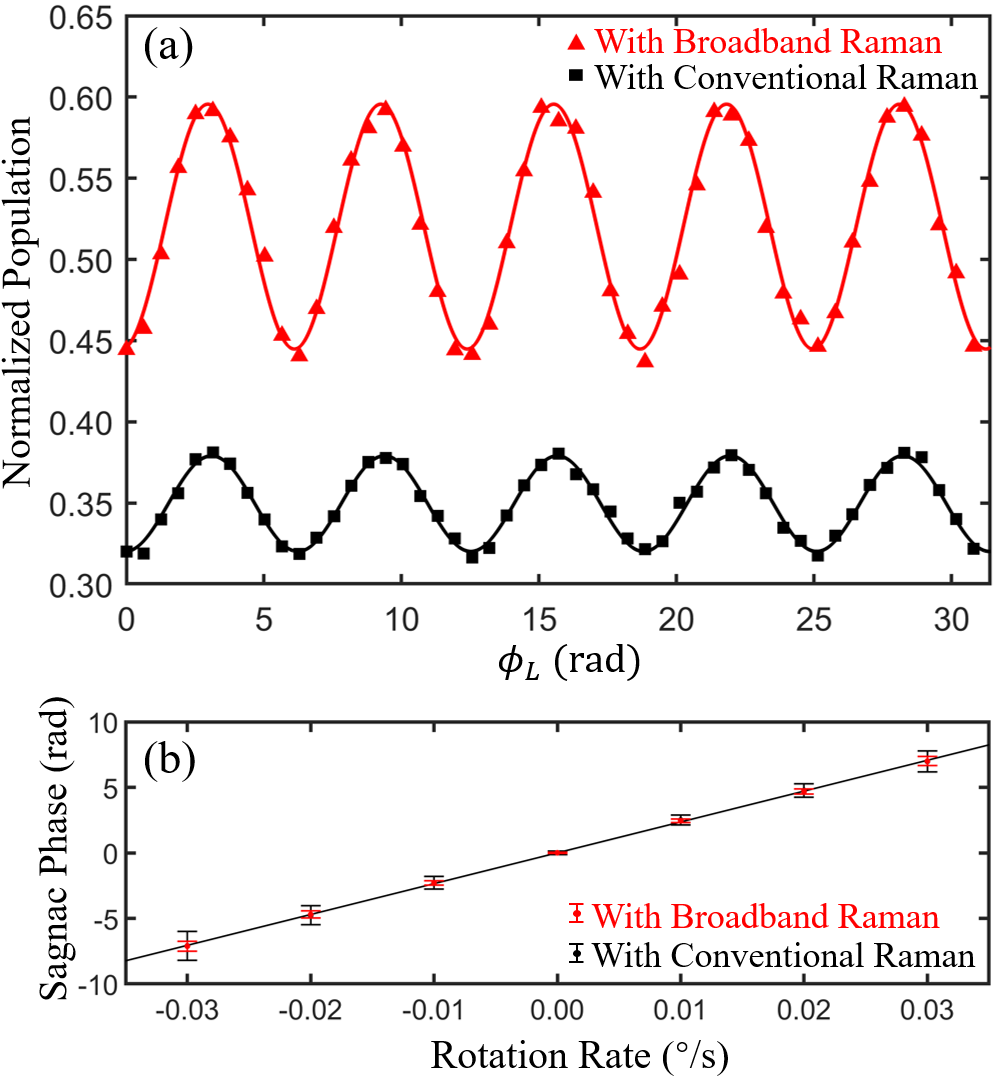}
\caption{\label{fig:fringe}Atom-interferometric performance with spectral-domain coherent control. (a) Interference fringes of the Mach–Zehnder atom interferometer obtained with conventional (black) and broadband (red) Raman pulses, measured by scanning the optical phase $\phi_{\rm L}$ of the final pulse. The broadband scheme yields a substantially increased fringe contrast, indicating a larger fraction of atoms participating in the interferometer. (b) Phase response of the interferometer under rotation, showing that the scale factor remains unchanged under spectral-domain coherent control.}
\end{figure}

We next evaluate the impact of spectral-domain coherent control on atom interferometry. A Mach–Zehnder interferometer is implemented using a $\pi/2$–$\pi$–$\pi/2$ pulse sequence. The interferometric phase is scanned by varying the optical phase $\phi_{\rm L}$ of the final pulse. As shown in Fig.~\ref{fig:fringe}(a), the broadband scheme leads to a substantial enhancement of the interference fringe contrast from $5.9(2)\%$ to $15.1(2)\%$. Since the contrast is defined as the fringe amplitude normalized to the total atomic flux, this improvement directly reflects an increased fraction of atoms coherently participating in the interferometer. This result establishes that spectral engineering of Raman transitions translates into a direct enhancement of interferometric performance by increasing effective atomic participation.

The broadband interferometer can be viewed as an ensemble of interferometers distributed in momentum space, each associated with a different two-photon detuning. Since these interferometers enclose identical areas, they share the same rotation sensitivity and contribute constructively to the total signal. This interpretation is confirmed experimentally by measuring the phase response under controlled rotation. As shown in Fig.~\ref{fig:fringe}(b), the phase responses of the broadband and conventional schemes coincide, indicating that the interferometer scale factor remains unchanged.

\begin{figure}[t]
\includegraphics[width=1.02\linewidth]{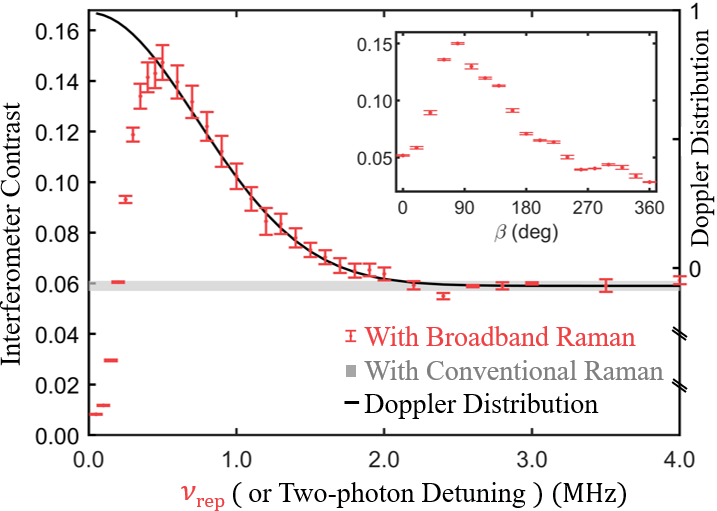}
\caption{\label{fig:Confrep}Dependence of interferometer performance on the spectral configuration of the broadband Raman coupling. Fringe contrast as a function of the frequency spacing $\nu_{\rm rep}$. The contrast exhibits a maximum at $\nu_{\rm rep}=500~\mathrm{kHz}$, reflecting optimal spectral matching between the Raman resonances and the Doppler distribution.}
\end{figure}

The performance of the broadband scheme is governed by the spectral relationship between the Raman resonances and the Doppler distribution. The spectral distribution of the resonances can be tuned by the modulation frequency $\nu_{\rm rep}$ and depth $\beta$. As shown in Fig.~\ref{fig:Confrep}(a), the interferometer contrast exhibits a clear maximum at $\nu_{\rm rep}=500~{\rm kHz}$, which is comparable to the transit-time-limited linewidth. This behavior reflects an optimal spectral matching between the Raman resonances and the atomic velocity distribution. At larger $\nu_{\rm rep}$, the spacing between adjacent resonances exceeds the Doppler width, so that fewer velocity classes are addressed. As a result, the contrast decreases and gradually approaches that of the conventional Raman transition. This trend follows the Doppler distribution and can be understood as a reduction in the number of atoms contributing to the interferometer until only the central ($n=0$) resonance remains effective. At smaller $\nu_{\rm rep}$, neighboring resonances overlap within the transit-time-limited linewidth, leading to crosstalk and reduced coherence. In the time domain, this corresponds to a modulation period longer than the interaction time, resulting in a time-dependent phase that induces dephasing.

In conclusion, we demonstrate that spectral-domain coherent control enables spectral shaping of stimulated Raman transitions, overcoming the mismatch between narrow Raman linewidths and Doppler-broadened atomic ensembles. By synthesizing multiple effective two-photon frequencies, the interaction addresses a wide range of atomic velocities, increasing effective atomic participation in atom interferometry. Experimentally, this approach enhances the fringe contrast of a Mach–Zehnder interferometer from $5.9(2)\%$ to $15.1(2)\%$ while preserving the interferometer scale factor. These results establish spectral-domain coherent control as a powerful paradigm for manipulating atom–light interactions in inhomogeneously broadened systems. In particular, the spectral-domain scheme breaks the  conventional trade-off between spectral bandwidth and beam geometry, enabling broadband coherent manipulation with simple, well-collimated Raman beams.

More generally, the spectral-domain approach is not restricted to Raman transitions but applies to a broad class of two-photon coherent processes subject to inhomogeneous detunings. Similar spectral-shaping strategies can be implemented in Bragg transitions \cite{Giltner1995,Ahlers2016,Pandey2026} and coherent population trapping \cite{Aspect1988,Balram2016,Ma2025}, where velocity or frequency selectivity plays a central role. The ability to tailor both amplitudes and phases of the frequency components further suggests a route toward quantum optimal control in the spectral domain, complementary to conventional temporal approaches. More broadly, spectral-domain coherent control provides a general framework for manipulating quantum systems affected by frequency errors.

The current implementation does not represent a fundamental limitation of the spectral-domain approach. The number and distribution of accessible Raman resonances can be further engineered by tailoring the modulation waveform applied to the rf signal \cite{Torres-Company2008,Wu2010}, enabling the generation of broader and more flexible spectral profiles to address higher-temperature atomic ensembles. Furthermore, the ability to selectively address different velocity classes suggests new possibilities for Doppler-resolved atom interferometry, analogous to spatiotemporal encoding in magnetic resonance imaging \cite{Tal2006,Qiu2024}. These perspectives highlight spectral-domain coherent control as a versatile platform for engineering atom–light interactions in inhomogeneously broadened quantum systems.


~

\emph{Acknowledgments}—The authors thank Dr. Nan Zhao (Beijing Computational Science Research Center) for helpful discussions on the theoretical calculation.



\bibliography{apssamp}

\end{document}